\documentstyle[sprocl2,epsfig]{article}
 
\input{psfig}
\def\Journal#1#2#3#4{{#1} {\bf #2}, #3 (#4)}
 
\def\NPB{{\em Nucl. Phys.} B}
\def\PLB{{\em Phys. Lett.}  B}
\def\PRL{\em Phys. Rev. Lett.}
\def\PRD{{\em Phys. Rev.} D}
\def\ZPC{{\em Z. Phys.} C}

\def\be{\begin{equation}}
\def\ee{\end{equation}}
\def\bea{\begin{eqnarray}}
\def\eea{\end{eqnarray}}

\begin{document}
 
\title{THERMAL GAUGE THEORY, EFFECTIVE ACTION AND Z(N) SYMMETRY }
 
\author{ S. Bronoff}
 
\address{Centre Physique Th\'eorique au CNRS, Case 907, Luminy\\ F13288,
Marseille Cedex, France}
 
\author{R. Buffa}
 
\address{Centre Physique Th\'eorique au CNRS, Case 907, Luminy\\ F13288,
Marseille Cedex, France}
\author{C.P. Korthals Altes}
 
\address{Centre Physique Th\'eorique au CNRS, Case 907, Luminy\\ F13288,
Marseille Cedex, France}
 
\maketitle\abstract{
One of the salient features of high temperature gluodynamics is the Z(N) groundstate degeneracy, key to the understanding of the transition. On the other hand impressive progress has been obtained with effective theories, that seemingly lack Z(N) symmetry. We give a simple unified description
of the two.  As a very useful byproduct we get a natural definition for the Debye screening mass. We show its expediency by calculating  the next to leading order in the
Debye mass. The sign of this correction gets a simple explanation. 

\section{Introduction.}{\label{sec:state}}
 
Thermal gauge theory is the main tool to study equilibrium properties
of plasmas in elementary particle physics. Their relevance for ongoing
experiments at CERN, for the physics at RHIC and LHC motivates
their study, both through analytic and numerical methods.

In recent years very useful results have been obtained through effective
thermal theories~\cite{laineqcd,laineew,rummukainenzn}. Properties of
the bulk plasma phase have been elucidated, leading to new insights.
Markedly absent in effective theories are the Z(N) symmetries, that gave
so much impetus to the understanding of the deconfining phase transition~\cite{svet}.
An inroad is made to the relation between these two aspects in ref.~\cite{rummukainenzn}.

 The present note gives a simple unified picture, and  some concrete consequences of this picture in perturbation theory~\footnote{ It is known for
some time that convergence is slow for certain observables~\cite{laineqcd,karschheller,kastening}. So its use for phenomenology is not clear. Nevertheless we feel that control of these aspects still warrants efforts.}.   


We  concentrate on pure SU(N) gluodynamics at some temperature
T, well above the critical temperature $T_c$, where deconfinement
takes place. Such a theory is characterized by the value of $\Lambda_{\bar min}$
and T. Their knowledge determines the coupling g. In perturbation theory
at high T (so $g=g(T)$ small)  three different scales come into play: T, gT , and $g^2T$. The electric screening length $m_D$ is of order gT, and there is a magnetic glueball mass of order $g^2T$ that limits the applicability of perturbation theory.

Now we consider a 3D box of size $L_{tr}^2L_z$, elongated in the z-direction,        containing the plasma.
Its statistical properties are described by the Gibbs sum over physical states:
\be
\exp{-F/T}=\sum_{phys}\exp{-H/T}
\label{eq:gibbs}
\ee
Physical means we impose Gauss' law everywhere inside the box.
Only at the sides of our box we may drop Gauss' law and impose 
special boundary conditions, specified below. The Hamiltonian
equals
\be
H=\int d\vec x Tr\left(g^2\vec E^2+{1\over g^2}\vec B^2\right)
\ee 
The sum can be rewritten as a Euclidean path integral with period $1/T$
in the time direction:
\be 
\exp{-F/T}=\int DA_0D\vec A\exp{-{1\over g^2} S(A)}
\label{eq:free}
\ee
The Z(N) invariance is naturally defined in terms of the Wilson line:

\be
P(A_0)\equiv {1\over N}Tr P\exp{i\int_0^{1/T}A_0(\tau,\vec x)}
\label{eq:wline}
\ee
Now the allowed gauge transformations are those that leave the action
invariant. Apart from periodic ones, we have transformations that are
periodic modulo the centergroup elements $\exp{ik{2\pi\over N}}$. Those are
the only ones leaving the action invariant. So the {\it{phases}} $C^i(\vec x)$ of the  eigenvalues of the loop are gauge invariant modulo multiples of $2\pi/N$ for all $i$ running from $1$ to $N$. What is important is that any adjoint field- or more generally any representation with zero N-ality- stays periodic (or anti-periodic) under any allowed transformation\footnote{Also for non-zero triality representation the action stays invariant. But the periodicity is affected}. 
 
Anticipating on the wall profile $p(z)$ in the z-direction we define the phase average $\bar C^i(z)={1\over{L_{tr}}^2}\int d^2x_{\bot} C^i(\vec x )$. 

The free energy of our system (~\ref{eq:free}) can be written as:
\be
\exp{-F/T}=\int Dp(z)\int DA_0 D\vec A\hspace{2mm}\delta(p(z)-\bar C)\exp{-{1\over g^2} S(A)}
\label{eq:intprofile}
\ee

So we split the path integration into an integration over profiles $p^i$
and the remaining variables (for readability's sake we left out the color
indices in eq.~\ref{eq:intprofile}. The remainder is an integration over all potentials with the gauge invariant delta function constraint. For very large size of the box we can write it as:

\be
\exp{-{f\over T} L_{tr}^2L_z-{1\over {g^2N}}{S_{eff}(p)\over T}L_{tr}^2}=
\int DA_0 D\vec A\hspace{2mm}\delta(p(z)-\bar C)\exp{-{1\over g^2} S(A)}
\label{eq:effaction}
\ee

The first term in the exponent is the free energy per unit volume
and does not depend on the profile $p$, which describes surface effects. 

The second term is the center of our interest. It contains the information on the dynamics of the wall. In the limit of very large box size $L_{tr}$
becomes so large that only the profile with minimal action will contribute.  So in that limit knowledge of the minimum of  $S_{eff}$  is
sufficient. Note that the effective action takes the same value for all p that differ by a centergroup transformation, because a gauge transformation periodic modulo a centergroup element will not change the value of the integral in eq.~\ref{eq:effaction}. Only the phase of the
Wilson line in the argument of the delta function will change by a shift $k{2\pi\over N}$. Boundary conditions on p at the edge of the box at $z=\pm{L_z\over 2}$ fix the minimum or domainwall as in fig. 1 and correspond to fixing the value of the Wilson line to be 1, respectively $\exp{i{2\pi\over N}}$.    

A last remark concerns the physical meaning of the profile, where it merges into the groundstate
of the plasma, on the far left or right in fig. \ref{fig:wall}. From its definition in eq.~\ref{eq:wline}  it is a gauge invariant version of the $A_0$ potential. In the Abelian case it is just the line integral of $A_0$, and we know  that it decays exponentially in the plasma. Hence we will take
\be
m_D\equiv lim_{z\rightarrow\pm\infty}\left\vert{{\partial_z p\over p}}\right\vert
\label{eq:debyedef}
\ee
as the definition of the Debye mass in the non-Abelian case. This definition is gauge invariant and non-perturbative. It is odd under Euclidean time reversal, so
cannot excite magnetic glueballs~\cite{arnoldyaffe}. We will see that it coincides in perturbation theory with the usual value of the Debye mass as defined by the two point function~\cite{rebhan} or by the correlation of the imaginary part of the Wilson line~\cite{arnoldyaffe,nadkarni}.

\section{Computing the effective action}{\label{sec:compu}}

In this section the strategy of our computation is laid out.
We suppose the coupling g is very small. Apart from the three scales
in the theory: $T$, $gT$ and $g^2T$ there is a scale due to the profile $p$, which is sliding.  As we saw in the previous section, $Z(N)$ invariance  introduces a periodicity in the profile, so $p/T=O(1)$ or less.

We follow the idea in ref.~\cite{ginsparg} and integrate out the heavy degrees of freedom first. Heavy means $O(T)$. Then we are left with an effective action which is  at most three dimensional, because the heavy degrees of freedom include all non-static modes.

We can distinguish two cases, depending on the value of $p/T$. 

\begin{itemize} 
\item[(i)] $p/T=O(1)$. Then we can integrate out all degrees of freedom and find $S_{eff}$ as a function of the profile p directly.
\item[(ii)] $p/T=O(g)$ or smaller. All non-static degrees of freedom are integrated out, but not the static ones. The resulting 3D action containing the static variables has a built in scale, the Debye mass. It is an action with non-local vertices. Only when we are looking at distances much larger than the Debye scale we can approximate the vertices by local ones. Hence we will treat the case $p/T=O(g)$ later, and discuss only the $O(g^2)$ case in this letter. 
\end{itemize}

\section{Perturbative saddle point and integration of heavy modes} 

In order to integrate out the heavy degrees of freedom we need perturbative information on the effective action, so we have to  search for the saddle point. This saddle point cannot be anything else than the a priori profile p appearing in the constraint. A simple analysis of the constraint shows this immediately, see below eq.~\ref{eq:constrainto2}. This means the fluctuations $Q$ around the profile can be written as $A_{\mu}=p\delta_{\mu,0}+gQ_{\mu}$, where p is a diagonal traceless $N\times N$ matrix, and Q any Hermitian traceless matrix. Then we have to fix the gauge, and the obvious choice is background gauge :
\be
{1\over{\xi}}Tr\left(D(p)_{\mu}Q_{\mu}\right)^2.
\ee
To deal effectively with the z-dependence of our background field $p(z)$ we will do a gradient expansion. This is justified, because the classical action of our profile is proportional to $1/g^2(\partial_z p)^2$. The next term is of quantum origin (the log of the fluctuation determinant) and is therefore O(1). Hence the minimum profile will have a gradient of order g to balance the two terms. 

So we will split the profile $p(z)=p+\delta p$. The first term is constant. The second term is expanded out and gives rise to vertices. Were it not for the constraint, the resulting Feynman rules would be the usual background field gauge rules (see e.g. ref.~\cite{abbott}) with all time derivatives $\partial_0$ replaced by their covariant counterparts $D_0(p)$, since p can be of the same order as $\partial_0$, that is, $O(T)$. So propagators for off-diagonal quantum fields change accordingly, and specially for the constant modes the profile acts like an infrared cut-off. Diagonal quantum fields do not feel the difference between the two derivatives, since p is diagonal, so commutes. So for SU(N) only N-1 massless excitations remain.

To see the consequence of the constraint in more detail we first expand the constraint in terms of the coupling g.

After some algebra we get 
\be
\bar C^i(z)-p^i(z)={1\over L^2_{tr}}\int d^2x_{\bot}\left({g\over 2}Q_0^i(z)+{g^2\over 2}\sum_{n,j\ne i}\left({1\over q^0_{ij}(p)}(Q_0^{ij}(\vec x,n)Q_0^{ji}(\vec x,-n)\right)+O(g^3)\right)
\label{eq:constrainto2}
\ee
in the traditional notation $Q^a=TrQ\lambda^a$, $Q^{ij}=TrQ\lambda^{ij}$,
with $\lambda^a$ the Gell-Mann matrices and $\lambda^{ij}$ the matrix with only one non-zero entry in the (ij)th place, all normalized to $1/2$.
The presence of the momentum component $q^0_{ij}(p)\equiv 2\pi Tn+p^i-p^j$
reflects the non-locality in time of the constraint.

So to lowest order the constraint tells us not to integrate over fluctuations that are diagonal and constant in the transverse directions.
This is to be expected, because they are zero modes of the profile $p(z)$. 
The only linear terms in the classical action in eq~\ref{eq:effaction} are
these same zero modes, so we have justified the profile $p(z)$ as the saddle point.

Now we expand the $O(g^2)$ terms out of the delta function. To reinstore
the delta function constraint we have to do a partial integration with respect to the zero modes. This is going to generate couplings of the $O(g^2)$
term of the constraint
to all derivatives of the action. These couplings stem from the non-linearity, i.e. the gauge invariance of the constraint and are therefore instrumental for the consistency of the approach.
 

In fig. 2 we show the lowest order couplings.
 The b couplings are renormalizing the kinetic
term and the q terms the potential terms in $S_{eff}$ much in the vain of 
Belyaev~\cite{belyaev}.

\section{Results of integrating heavy modes}{\label{eq:heavyint}}
We now integrate the right hand side of eq.~\ref{eq:effaction} over the non-static
modes~\cite{ginsparg}. The result will be of the form:
\be
\exp\left(-{1\over {g^2N}}{S_{eff}(p)\over T}L_{tr}^2\right)=\int DQ(0)\exp\left({-{S^{(3)}(Q(0),p)\over T}-{1\over {g^{2}N}}{{\hat S_{eff}}\over T}L_{tr}^2}\right)
\label{eq:neq0result}
\ee
  
In the limit of small external momenta the three dimensional action $S^{(3)}$ can be written in
a local form:

\bea
S^{(3)}(Q,p)=\int d\vec x& &1/2\sum_{k\neq l}Tr\left(D(Q)_kQ_l\right)^2+Tr\left(D(Q_k)Q_0\right)^2+Tr\left([ip,Q_{\mu}]^2\right)\nonumber \\ &+&{1\over{\xi}}Tr\left(\partial_kQ_k\right)^2+m_D^2(p)TrQ_0^2+ \delta S^{(3)}
\label{eq:3Daction}
\eea

With the profile p set to zero it reduces to the effective action used in
computations concerning observables in the groundstate~\footnote{Usually the gauge fixing term in the static sector and hence ghost terms are dropped: the static sector is considered as a non-perturbative sector. Here, like in \cite{rebhan,braatenpolyakov} we keep
the gauge fixing to pull out precisely the interesting-and still available-perturbative information from the static sector.}.~\cite{laineqcd,rummukainenzn} 

The coefficients  in $S^{(3)}$ are of order $g^2$ (Debye mass) or $g^4$ ($\delta S^{(3)}$) and stay so when p is turned on~\cite{bronoffetal}. For the next to leading order in the Debye mass we will only need the coefficient $m^2_D(0)={N\over 3}g^2T^2$. 
The remaining terms in eq.~\ref{eq:neq0result} are the truncated kinetic and potential terms in $S_{eff}(p)$.

\subsection{Profile of order $T$}

The p-induced mass terms are eventually getting larger than the Debye mass:
in case the profile $p/T$ is of order one we can integrate out the static modes
on the same basis  as the non-static modes%
, and  change  $\hat S_{eff}(p)$ to $S_{eff}\equiv K+ V$. K is the kinetic term and V the potential.
Since these has been analyzed including two loop order we shall be brief and refer the reader for more details to the original
papers~\cite{bhatta,cpka,enquist}.
One obtains for all those values of p a perfectly Z(N) invariant profile
(see fig. \ref{fig:wall}).

\begin{figure}[ht]
\centering
\epsfig{figure=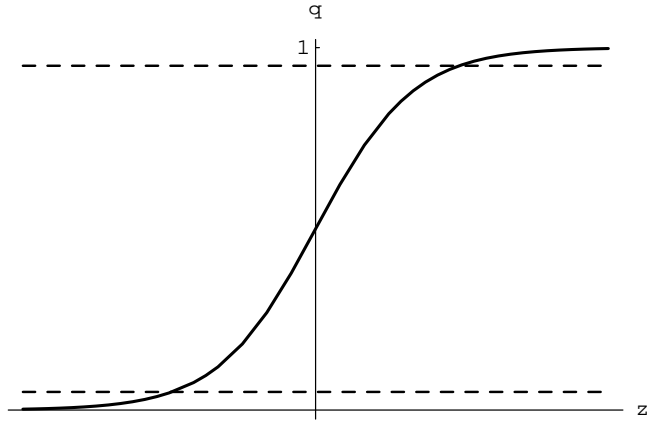,width=8.5cm,angle=0}
\caption{Profile of the wall. Dashed horizontal lines separate the $q=O(1)$ region from $q=O(g)$.}
\label{fig:wall}
\end{figure}

We give the explicit expression for SU(N). We limit ourselves to the valley of minimal
action. It is parameterized by one parameter q, that determines the matrix p as:
\be
p=2\pi Tq\hspace{2mm}diag (1/N,1/N,....,(1-N)/N)
\ee
The potential part V reads:
\be
V(q)=(N-1)(2\pi T)^2m_D(0)^2\left(1-5{g^2(T)N\over{(4\pi)^2}}\right)\int dz q^2(1-q)^2
\label{eq:potresult}
\ee
with $m_D(0)$ the perturbative one loop Debye mass in eq.~\ref{eq:3Daction} and q
mod 1.
The first term comes from the fluctuation  determinant.

 The second term comes from
the two loop free energy graphs and from the renormalization of the Wilsonline through the q-vertex in fig. \ref{fig:vertex}. The latter is a typical example of how the vertices from the constraint restore gauge independence\cite{belyaev,cpka}. 

\vspace{1cm}
\begin{figure}[ht]
\centering
\epsfig{figure=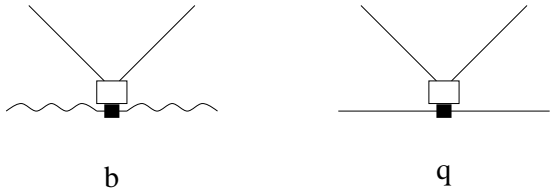,width=2.5cm,angle=-90}
\caption{Lowest order additional vertices coming from the constraint. The lines attached to the empty square are the Q's in eq. \ref{eq:constrainto2}. The lines attached to the black square are due to differentiation of the three-vertex of the invariant action as explained below eq. \ref{eq:constrainto2}. Vertex $q$ is $O(g^2)$, vertex $b$ $O(1)$.}
\label{fig:vertex}
\end{figure}

The result for the kinetic term reads:
\be
K=(N-1)(2\pi T)^2\int dz\left(1+{11\over 3}{g^2(T)N\over{(4\pi)^2}}\left(\psi(q)+\psi(1-q)+1\right)\right)(\partial_z q)^2
\label{eq:kinresult}
\ee

The first term is the classical term. Combining this term with the lowest order term in V we find indeed that for small values of the profile the perturbative Debye mass controls its behavior for large  z. So our definition of the Debye mass at the end of section~\ref{sec:state} is justified.

The second term is obtained from the graphs in fig. \ref{fig:kinetic}.
As is well known, these graphs give at zero temperature a p-independent result (at zero
temperature we deal with an integration instead of summation over frequencies, so all p dependence is absorbed). All what 
stays is coupling constant renormalization.

At finite temperature we are left with q dependence through the digamma functions. Gauge dependence in these finite parts is 
absorbed by the b-vertex induced contribution from fig. \ref{fig:vertex}.
Since we integrated out the fluctuations at scale T the typical coupling constant renormalization coefficients are expected.  The renormalized coupling in the
results is obtained by dimensional regularization, and defined by:
\be
g^{-2}(T)=g^{-2}(\mu)\left(1-{11\over 3}{g^2(\mu)N\over{(4\pi)^2}}\left({1\over{\epsilon}}-\log\left({\pi T^2
\over{\mu^2}}\right)+\psi(1/2)\right)\right)
\label{eq:rencoupling}
\ee

Some  comments are  in order.
The results for K and V are the leading terms in a gradient expansion, where the gradient is units of the temperature T or the profile p.
 
Trying to obtain the next to leading order for the Debye mass from eq.~\ref{eq:kinresult} leads to disaster, since
the digamma function behaves like ${-1\over q}$ at small q. It is due to the static contribution, and an artifact
of our use of Feynman rules not including the effective 3D action with the Debye mass. The latter is
only justified when $q=O(1)$.


\subsection{The profile for $p=O(g^2)$, and corrections to the Debye mass}

As we just saw the  kinetic and potential terms are in this region dominated by the static term in K.
We know from Linde's veto~\cite{linde} that we cannot take the cut-off $p/T$ smaller than $g^2$, otherwise perturbation theory makes no sense anymore. 
From the explicit form for K we see that we will get the leading contribution for these values  of p.
So we computed the first four graphs given in fig. \ref{fig:kinetic} with the rules from $S^{(3)}$. In this action propagators involving $Q_0$ have now the Debye mass
{\it{ dominating}}  the mass term induced by the profile. In 
the propagators
where the Debye mass is absent the profile still provides a cut-off $2\pi Tq\equiv m_s$. This renders our approach unambiguous.
We have calculated with these rules the four diagrams in fig. \ref{fig:kinetic}
and for any value of the gauge parameter $\xi$. The leading  result comes entirely from the first graph in fig. \ref{fig:kinetic}. It is for all $\xi$ equal to:
\be
2{1\over{(4\pi)^2}}4r^2 {2\pi T\over{ir}}\log\left\vert{ir+m_s+m_D(0)\over{-ir+m_s+m_D(0)}}\right\vert
\label{eq:leading}
\ee
\begin{figure}[ht]
\centering
\epsfig{figure=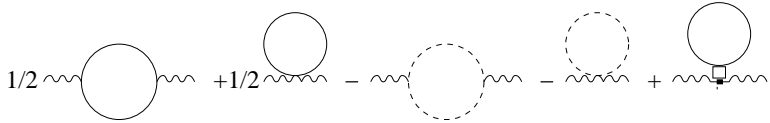,width=1.6cm,angle=-90}
\caption{1-loop corrections to the kinetic term. Dashed lines are ghost lines, straight lines  Q-lines, wavy lines background fields. The last graph on the right comes from the vertex b in fig. \ref{fig:vertex}.}
\label{fig:kinetic}
\end{figure}

The  logarithmic term gives the cuts starting at $\pm i(m_s+m_D(0))$ in the momentum r of the incoming line. Since the momentum r equals $im_D(0)$ the contribution to the kinetic term becomes:
\be
K=(N-1)(2\pi T)^2\left(1-2{g^2N\over{4\pi}}{T\over{m_D(0)}}\log\left({m_D(0)\over {m_s}}\right)\right)(\partial_z q)^2
\ee

Plugging this result into the equation of motion we get as correction 
to the Debye mass $m_D$:
\be
m_D=m_D(0)\left(1+{g^2N\over 4\pi}{T\over m_D(0)}\log\left({m_D(0)\over m_s}\right)\right)
\label{eq:finalresult}
\ee
To leading order the argument of the log can taken to be $1/g$. This is the result of Rebhan~\cite{rebhan}. 
\section{Conclusions}
We have given a unified treatment of the Z(N) invariant action and the 
effective action. In effective action parlance: in between two Z(N) vacua
constant modes get heavy too, and can be integrated out with out recourse
to non-perturbative physics. Near the Z(N) vacua the constant modes get
light, and if they are light enough- $O(g^2 T)$ (but not lighter)- one can use the 3D effective
action to compute the perturbative effects. 

Our approach brought two boons: a manifestly gauge invariant and non-perturbative definition
of the Debye mass, and a gauge invariant infrared regulator. It warrants the computation of higher order perturbative effects in the surface tension, and in the bulk free energy at $O(g^6)$. 

The sign of the correction in eq.~\ref{eq:finalresult} is interesting. It 
is correlated to the sign of the coupling constant renormalization. At scales
T  it is opposite in Abelian and non-Abelian theories. 
This sign difference still persists  at scales $g^2T$, where we evaluate the
Debye mass. Only the magnitude changes, both numerically and in the order of the 
coupling. This correlation is manifest in our choice of observable.

An important issue is whether the  $g\log(1/g)$ correction is not reproduced in
 higher order, admitting higher order vertices in the effective action $S^{(3)}$. That is, one hopes to have at most $O(g)$ corrections. In the context
of the two-point function definition one has never found any counter examples.
With the definition advocated above it may be easier to find a truly satisfying argument. While the non perturbative $O(g)$ corrections dominate  the  $g\log(1/g)$ corrections till very high temperature~\cite{laineqcd}, the center of the domainwall remains accessible to perturbation theory at relatively low $T$.


It is an open question whether there exists a  version of our definition of the Debye mass that would permit us to study on a 3D lattice the wings of the domainwall. Breaking $Z(N)$ symmetry by adding a complex representation does not affect our method of calculating bulk properties. 
\section*{Acknowledgements}
We are indebted to Luigi del Debbio for help in the early stages of this work, and K.Farakos, K. Kajantie, B. Kastening, M. Laine, A. Rebhan and M. Shaposhnikov for useful discussions. Thanks are due to M. Shaposhnikov and A. Rebhan for reading the manuscript.
S.B and R.B. thank the M.E.N.E.S.R. for financial support.

\end{document}